\documentclass[a4paper]{article}

\usepackage{INTERSPEECH2019}
\usepackage[numbers,sort&compress]{natbib}
\usepackage{booktabs,url}
\usepackage{xcolor}

\newcommand{\vect}[1]{\mathbf{#1}}

\title{Multi-Span Acoustic Modelling using Raw Waveform Signals}
\name{{P. von Platen$^{1,2}$, C. Zhang$^1$, P. C. Woodland$^1$}}

\usepackage{comment}
\address{$^1$ Cambridge University Engineering Dept., Trumpington St., Cambridge, CB2 1PZ U.K. \\
	 $^2$ Institute of Communication Systems (IKS), RWTH Aachen University, Germany
}
\email{\small \tt \{pwv20,cz277,pcw\}@eng.cam.ac.uk}

\begin{document}

\maketitle
\begin{abstract}
Traditional automatic speech recognition (ASR) systems often use an acoustic model (AM) 
built on handcrafted acoustic features, such as log Mel-filter bank (FBANK) values.
Recent studies found that AMs with convolutional neural networks (CNNs) can directly use
the raw waveform signal as input.
Given sufficient training data, these AMs can yield a competitive word error rate (WER) to those 
built on FBANK features.
This paper proposes a novel multi-span structure for acoustic modelling based on the raw waveform
with multiple streams of CNN input layers, each processing a different span of the raw waveform signal.
Evaluation on both the single channel CHiME4 and AMI data sets show that multi-span AMs give a lower WER than FBANK AMs 
by an average of about 5\% (relative). 
Analysis of the trained multi-span model reveals that the CNNs can learn filters that are rather different to the log Mel-filters.
Furthermore, the paper shows that a widely used single span raw waveform AM can be improved 
by using a smaller CNN kernel size and increased stride to yield improved WERs.

\end{abstract}

\section{Introduction}

Automatic speech recognition (ASR) systems usually consist of an acoustic model (AM) that captures the acoustic and phonetic properties of the speech signal and a language model (LM) providing
linguistic and syntactic context information at the word-level. 
Traditional AMs are normally built on handcrafted acoustic features, 
such as  log Mel-filter bank values (FBANK) or their approximate linear decorrelations known as
Mel frequency cepstral coefficients (MFCCs) \cite{davis1980comparison}.
These handcrafted acoustic features are broadly based on models from human speech production and 
perception \cite{hermansky1990perceptual,von1960experiments}  so that
they are not optimised toward the training criterion of the AM and might thus
discard valuable information from the raw waveform signal.

For AMs based on hidden Markov models (HMMs) with diagonal Gaussian mixture output distributions,
a compact feature representation such as MFCCs was required \cite{mitra2017robust}.
However with the resurgence of artificial neural networks (ANNs), along with increasing computational power,
there are far fewer restrictions  on the input  features, and using the raw waveform signal now becomes an interesting alternative 
to handcrafted acoustic features \cite{tuske2014acoustic,sainath2015learning}. 
AMs built on the raw waveform signal input make no prior assumptions about the data, which allows the AM to learn the most suitable raw waveform feature representation given sufficient training data.
Active research work has been carried out for the use of raw waveform features for 
acoustic modelling since 2014 \cite{sainath2015learning,ghahremani2016acoustic,tuske2018acoustic}, 
and has yielded competitive word error rates (WERs) to the standard approach using MFCC or FBANK features.
In \cite{sainath2015learning}, a 35ms window of the raw waveform signal is fed into 
a convolutional neural network (CNN) layer with rectified linear unit (ReLU) \cite{nair2010rectified} activation for 
time-frequency decomposition, followed by max-pooling and logarithm layers to imitate the 
logarithm compression of FBANK features. \\
Analogous to a frame, it produces a feature vector which is fed into a second CNN layer
\cite{cldnnGoogle}, similar to the AMs applying a frequency convolution over FBANK features \cite{Abdel-Hamid2014}.
In \cite{tuske2018acoustic}, the first CNN layer also performs a temporal convolution while the second CNN layer extracts the 
spectral envelope followed by logarithm or root compression \cite{hermansky1990perceptual}. 
Seventeen consecutive output vectors from the second CNN layer are then stacked to have a total input span of $291$ms, 
and the resulting output vector is fed into a deep neural network (DNN) with 12 fully connected layers. 
Non-linearities other than max-pooling with more discriminative kernels can be used 
to aggregate the output of the CNN input layer \cite{ghahremani2016acoustic}.
Zhu \textit{et al.} \cite{zhu2016learning} proposed another structure
in which CNN layers with different kernel sizes are configured to learn features of 
different time-frequency resolutions within a 20ms window, similar to wavelets \cite{haar1910theorie}.
Several other studies have investigated the use of raw waveform signal input from multiple 
microphones in far-field ASR \cite{sainath2015speaker,kim2017end}.
Analysis of the trained CNN layers with raw waveform input reveals a strong similarity between the learned kernels and audiological 
distributed narrow band pass filters such as log-Mel filter banks \cite{sainath2015learning,ghahremani2016acoustic,golik2015convolutional}. This finding has reaffirmed the effectiveness of using handcrafted acoustic feature inputs and has inspired joint training of only some of the feature extraction pipeline with the AM \cite{variani2016complex,zhang2017thesis,ghahremani2018acoustic}. However, it also motivates trying to learn feature representations that are different to handcrafted acoustic features, \textit{e.g.} \cite{zhu2016learning}. 
In this paper, we propose a  novel multi-span AM structure which combines multiple input streams to learn more diverse feature 
representations from different spans of the same raw waveform input. 
Each stream uses a stack of two consecutive CNN layers and each span is configured using the same kernel size but 
different stride numbers for temporal convolutions. 
Single channel experimental results on far-field CHiME4 data show that a 5 layer DNN with three streams 
outperformed the FBANK AM.
It can be observed that the learned filters are rather different to the log-Mel ones.
It may also noted that a set of small CNN kernels each having just 50 trainable parameters 
outperforms the set of larger CNN kernels each having 400 trainable parameters normally used for raw waveform input \cite{tuske2014acoustic,tuske2018acoustic,sainath2015speaker,golik2015convolutional}.
These findings are validated by experiments with data from headset microphones from the AMI data set.
The paper is structured as follows. \\
In Sec. 2, CNNs are revisited for raw waveform signal input. Section 3 explains in detail the proposed multi-span AM structure.
The experimental setup and results on CHiME4  and AMI are given in Sec. 4 and Sec. 5, with discussion in Sec. 6, 
followed 
by conclusions.

\section{Revisiting CNNs with Waveform Input}
CNNs \cite{lecun1998gradient} are powerful ANN models that can learn complex feature representations, as 
has been shown in image recognition with raw pixel input \cite{lecun1998gradient,vgg}.
Excluding the bias for simplicity, a CNN layer consists of $K$ trainable kernels, 
$\vect{w}_1,\vect{w}_2,\dots,\vect{w}_K$. 
Each kernel $\vect{w}_k$ is convolved over $T$ input samples of the raw waveform signal 
$\vect{x}_1^T$ with a  stride $S$ (denoted by $\ast^S$):
\begin{equation} \label{eq:cnn}
	\widetilde{\vect{y}}_k = \vect{w}_k \ast^S \vect{x}_1^{T} 
\end{equation}
where $\widetilde{\vect{y}}_k$ denotes the $k$-th (one dimensional) output feature map. The output from a CNN layer at 
each time step comprises of $K$ output feature maps, and the size of each map $M$ can be determined by
\begin{equation} \label{eq:inputOutput0}
	M=\lfloor(T-L)/S\rfloor+1,
\end{equation}
where $L$ is the kernel size and $S$ the stride.

Splitting the raw waveform $\vect{x}_1^T$ into $M$ overlapping 
windows $\left[x_1,\dots, x_L \right], \dots, \left[x_{1+SM}, \dots, x_{L+SM} \right]$, with $x_{j}$ representing the $j$-th sample of $\vect{x}_1^T$, then
\begin{equation}
	{\vect{y}}_m =
	\left[x_{1+S(m-1)}, \dots, x_{L+S(m-1)} \right]
	\left[ \vect{w}_1, \dots, \vect{w}_K \right],
\end{equation}
results in a vector ${\vect{y}}_m$ based on a fixed window of raw waveform using $K$ kernels. 
${\vect{y}}_m$ can be viewed as a ``frame'' similar to the one used in traditional acoustic feature analysis and can be obtained by extracting the 
$m$-th elements from all $K$ output feature maps.

\begin{figure}[h]
  \centering
  \includegraphics[width=\linewidth]{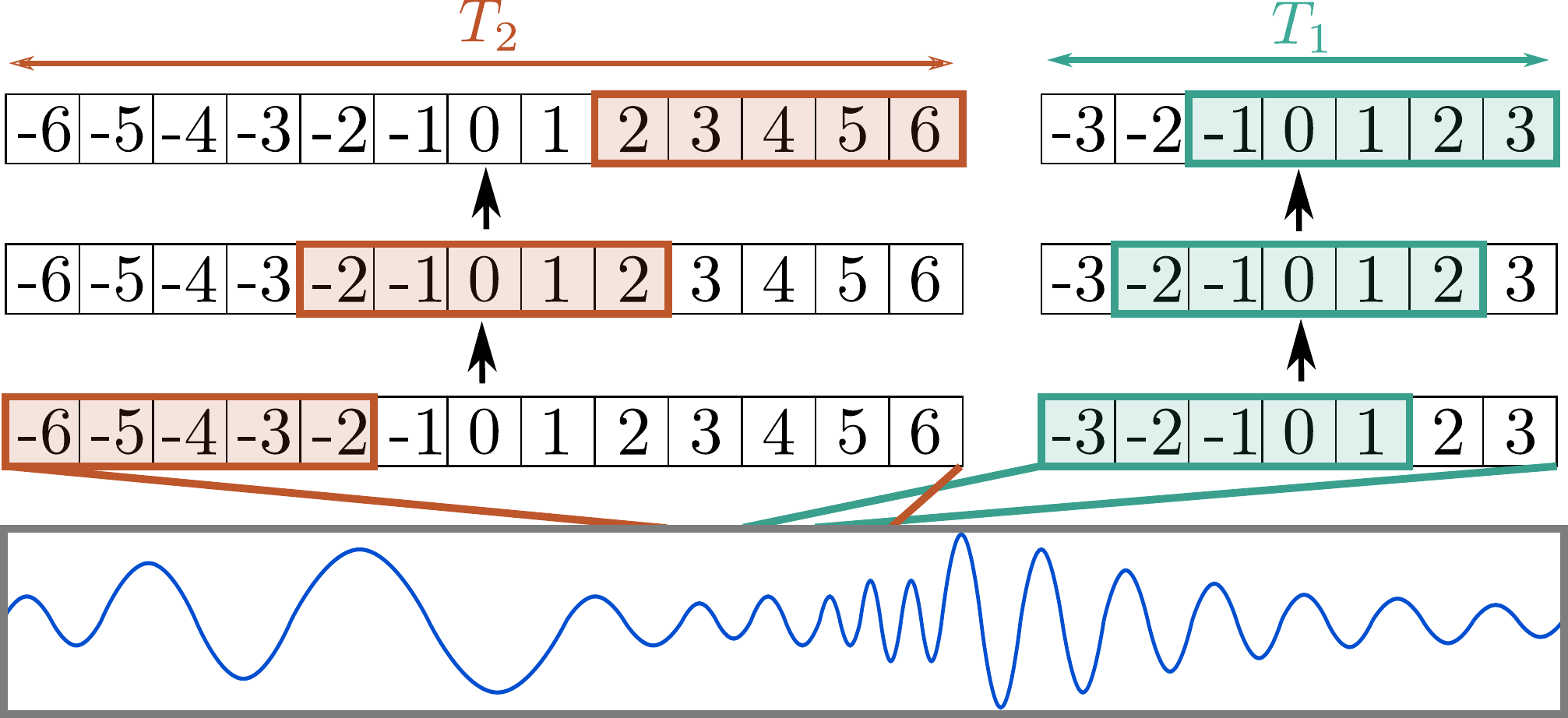}
  \caption{Examples for temporal convolution with a output feature map size $M=3$ and kernel length $L=5$. 
  The strides $S_1=1$ (green) and $S_2=4$ (orange) define the spans $T_1=7$ (green) and $T_2=13$ (orange) respectively.}
  \label{fig:conv}
\end{figure}

Two examples of CNN kernels of the same size $L=5$, but different strides 
$S_1=1$, $S_2=4$ and input spans $T_1=7, T_2=13$ are given in Fig.~\ref{fig:conv}.
From the figure and based on Eqn.~\eqref{eq:inputOutput0}, it is clear that the input span $T$ can be 
viewed as a function of $S$, $L$, and $M$, \textit{i.e.}
\begin{equation} \label{eq:inputOutput}
	T=(M-1)S+L.
\end{equation}
Therefore $T$ is controlled by varying $S$ while fixing $L$ and $M$. 
For example in Fig.~\ref{fig:conv}, both the orange and green kernels have the same size $L=5$ and 
yield an output feature map sized $M=3$, whereas the orange kernel considers 
a much larger input span of $T_2=13$ due to its bigger stride $4$.
In the rest of the paper, ${\vect{y}}_m$ will denote the $m$-th output feature vector.
Throughout the paper, the notation 
\begin{equation} \label{y}
	\vect{y}=\text{CNN}^{L}_{S}(\vect{x}^{T}_1,M)
\end{equation} 
is defined to denote a CNN layer, where $\vect{y} = \left[{\vect{y}}_1, {\vect{y}}_2,\ldots ,{\vect{y}}_M\right]$ is the concatenation of all $M$ output feature vectors.

\section{Multi-Span Acoustic Model}

Frames of traditional acoustic features, such as MFCC and FBANK, are usually derived using the short-time Fourier transform (STFT) 
based on a 25ms window, within which the speech signal is assumed to be stationary, and a window shift of 10ms. Conventional cross-entropy (CE) trained feed-forward DNN AMs have been found to yield the lowest WERs 
when 11 concatenated frames (or alternatively 9 concatenated frames if first order differentials are included) are used as the AM input 
\cite{mohamed2012understanding,Siniscalchi2013,Woodland2015}, 
which results in an input span of 125ms of the raw waveform signal.
Actually, it has been found that more powerful ANN AMs, such as recurrent or time-delayed neural networks, 
can effectively use a much longer span than DNNs \cite{waibel1989tdnn,Robinson:1996ab}.
This shows the importance of input span for acoustic modelling.

\begin{figure}[h!]
  \centering
  \includegraphics[width=\linewidth]{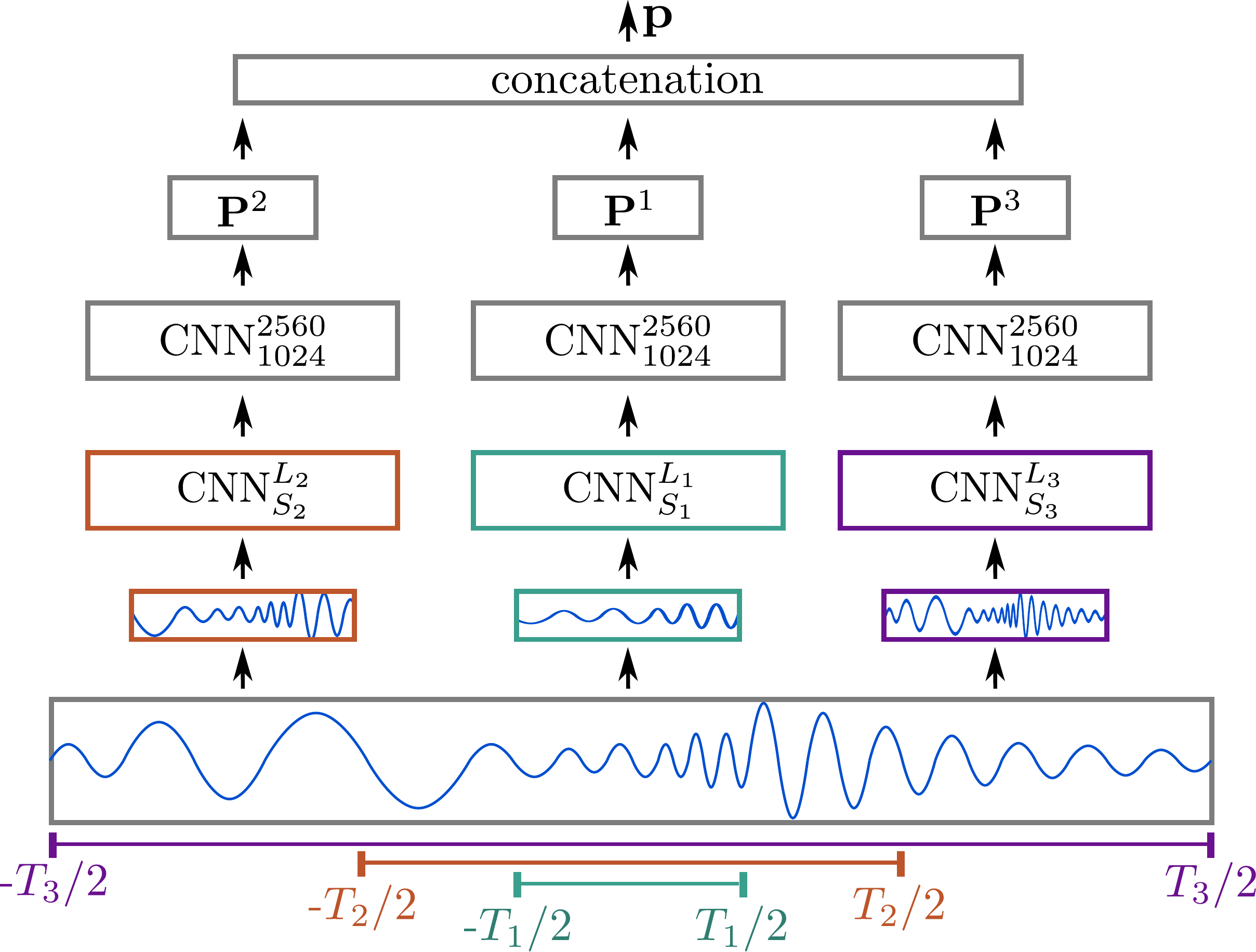}
  \caption{A sketch map of using three CNN input streams to convolve over different raw waveform spans
  based on the ranges of $[\text{-}{T_1}/{2},{T_1}/{2}],[\text{-}{T_2}/{2},{T_2}/{2}],[\text{-}{T_3}/{2},{T_3}/{2}]$ respectively.}
  \label{fig:model}
\end{figure}
\vspace{-4mm}

The multi-span AM is proposed in this paper, which improves FBANK based AMs by 
using multiple input streams to extract a more diverse set of complementary features 
from the raw waveform. 
As an example, three input streams of the multi-span AM are shown in Fig.~\ref{fig:model},
which produce the outputs $\vect{o}^{1}$, $\vect{o}^{2}$ and $\vect{o}^{3}$ from different 
spans $T_1$, $T_2$ and $T_3$ respectively by using two consecutive CNN layers.
More specifically, for each input stream $i$, 
CNN input layers are convolved over a unique span of the raw waveform signal $\vect{x}$ yielding
\begin{equation}
	\label{eq:cnn1}
\vect{y}^i=\text{CNN}^{L_{i}}_{S_{i}}(\vect{x}_{\text{-}{T_i}/{2}}^{{T_i}/{2}},M_i),
\end{equation}
where $S_{i}$, $L_{i}$, and ${M_{i}}$ are parameters defining the first CNN layer.
Next, $\vect{y}^i$ - which is a flattened array of length $M_iK$ (\textit{cf.} with Eq. \eqref{y}) - 
is fed into a separate second CNN layer convolutions with stride, kernel size, and output feature map size set to $S_{i2}$, $L_{i2}$, and $M_{i2}$, respectively, \textit{i.e.}
\begin{equation}
	\label{eq:cnn2}
	\vect{o}^i=\text{CNN}^{L_{i2}}_{S_{i2}}(\vect{y}^i,M_{i2}).
\end{equation}
Multiple CNN layers could be stacked in each stream which can result in the use of smaller kernel sizes \cite{vgg}.
The size of the resulting output $\vect{o}^i$ from each stream $i$ can be reduced by using a linear projection $\mathbf{P}^i$, and the 
 final multi-span feature vector $\mathbf{p}$ can be formed by concatenating $\mathbf{P}^i\vect{o}^i$ from all streams.

In this paper, only input streams with two CNN layers are investigated. 
For the CNN input layers given in Eqn.~\eqref{eq:cnn1}, $M_i=200$ and $K=64$ kernels are fixed throughout the paper,  
while for the second CNN layers, $M_{i2}=11$, $S_{i2}=1024$, and $L_{i2}=2560$ are used in this paper. 
The ReLU activation function is applied to the output of both CNN layers in each stream.
By fixing the kernel number of the second CNN layers to be 128, the size of each output $\vect{o}^i$ is  $128\times11=1408$, 
which is reduced to 150-d by $\vect{P}^i$.

It is to be emphasised that the only parameters that differ in each stream $i$ are the stride $S_i$ and the kernel size $L_i$ 
of the input CNN layers.
If the vectors $\vect{o}^i$ from all streams are of equal size, 
then the input span $T_i$ of the raw waveform signal for each stream is given by Eqn.~\eqref{eq:inputOutput}.

It is worth noting that in contrast to other models \cite{ghahremani2016acoustic, tuske2018acoustic,sainath2015speaker, hoshen2015speech}, 
there is no log-compression, root-compression, max-pooling or other special non-linearity used in our current setup in order to constrain the model as little as possible to learn 
the best possible feature representations from multiple input spans. 
It may be possible to further improve the multi-span model by \textit{e.g.} using different non-linearities 
for different input streams.

\section{Experimental Setup}

The proposed multi-span AM was evaluated by training systems 
on CHiME4 \cite{vincent2017analysis} and AMI \cite{carletta2005ami} 
using HTK 3.5.1 and PyHTK \cite{Young:2015ab,Zhang:2019ab}.
In the results reported here, the multi-span feature vector $\vect{p}$ of the concatenated input streams is fed into 
a simple feed forward DNN with $4$ hidden layers each 
having $512$ output nodes and ReLU activation function. The DNN output layer dimension corresponds 
to the number of clustered triphone-states and applies the softmax activation function.
This structure is abbreviated as \emph{4L-512d-DNN}.
We used rather small AMs without many parameters 
compared to other AMs using the same data sets \cite{menne2016rwth, menne18:lafrw}, to ensure a quick turn around.

The training data is aligned at 10ms frame intervals to the clustered triphone-states.
For both corpora, $10$\% of the aligned training data was held back for cross-validation.
All models were trained by the CE criterion, using stochastic gradient descent
optimization with momentum, weight decay and the 
$\text{NewBob}^{+}$ learning rate scheduler \cite{zhang2017thesis}. 
To match the number of alignment frames, the raw waveform input is 
shifted by 10ms or 160 samples after every forward pass of the model. 

\subsection{CHiME4}

Initial DNN AMs were trained 
on 18h of the training corpus recorded by a close talking microphone (tr05-org + channel 0 on tr05-real) 
and the alignments obtained were used for all subsequent experiments. The data was aligned 
at a 10ms frame interval level to one of 3006 clustered triphone-states.
The 18h training set for DNN AMs consisted of real and simulated data from channel 5.
The raw waveform signal input was globally normalised for both zero mean and unit variance.
Because of the known microphone failures
\cite{vincent2017analysis}, for every utterance, the channel 
used for decoding the 5.6h development (dev) set was chosen 
according to a microphone failure detection algorithm presented 
in \cite{menne2016rwth}. 
Speech recognition experiments were conducted using Viterbi decoding based on a 5k vocabulary 3-gram (tg) LM trained on the official CHiME4 LM training data.

\subsection{AMI}
The training data for AMI includes 78.2h of speech from individual headset microphones
(AMI-IHM). The alignments were generated based on
10ms frames and the decision trees with 3996 clustered triphone-states. 
Both FBANK and raw waveform data was normalised at the utterance level for zero mean and at the 
meeting level for unit variance.
The systems were evaluated with the official dev and evaluation (eval) 
sets, which contain 9.0h and 8.7h speech, using the official testing dictionary with an 49.4k word vocabulary \cite{carletta2005ami}, a 4-gram (fg) LM, and Viterbi decoding.

\section{Experiments}

Initially all systems were evaluated on the CHiME4 dataset.
At a later stage, key results were validated on the AMI dataset.

\subsection{CHiME4 Channel 5}%
\label{sub:chime4}

%\subsubsection{Baselines}%
%\label{sub:baselines}
The 4L-512d-DNN baseline based on the FBANK features is denoted as $F_{160}^{400}$ with 
$160$ and $400$ defining the filter shift and 
filter size in number of samples used in the STFT respectively{\footnote{In comparison with 
\cite{vincent2017analysis} where the AM is much larger, or \cite{weng2014recurrent} where the AM
uses recurrent layers and discriminative sequence training, 
the baseline $F_{160}^{400}$ WER in Table \ref{tab:single_span_chime} is reasonably good.}}. 
For the single-span AM using raw waveform signal input, the output $\vect{o}$ of a
single input stream 
\begin{equation}
	\vect{o}=\text{CNN}^{2560}_{1024}(\text{CNN}^{L}_{S}(\vect{x}_{\text{-}T/2}^{T/2}))
\end{equation}
is directly fed into a 4L-512d-DNN without dimension reduction.
We denote the single-span AM as $I_S^L$ with $L$ and $S$ 
corresponding to the kernel size and stride of the CNN input layer.
All weights were randomly initialised without any pretraining.

\vspace{-2mm}
\begin{table}[th]
  \centering
  \caption{\%WERs with a tg LM and AMs with single input stream on CHiME4 dev set.
  Stride $S$ and kernel size $L$ are varied, and $L$ and span length $T$ are counted in waveform samples and ms. }
  \begin{tabular}{ l c c c c c}
    \toprule
    \multicolumn{1}{l}{ID} & $S$ & $L$ & $T$ & dev  \\
    \midrule
    F$_{160}^{400}$ & 160 & 400 & 125 & 18.1 \\
    \cmidrule{1-5}
    $I_{10}^{400}$ & $10$ & $400$ & $149$ & 20.2  \\
    $I_{10}^{100}$ & $10$ & $100$ & $131$ & 19.4  \\
    $I_{10}^{50}$ & $10$ & $50$ & $128$ & 19.3    \\
    $I_{10}^{25}$ & $10$ & $25$ & $125$ & 20.7  \\
    \cmidrule{1-5}
    $I_{4}^{50}$ & $4$ & $50$ & 53 & 23.2  \\
    $I_{9}^{50}$ & $9$ & $50$ & 115 & 19.7  \\
    $I_{15}^{50}$ & $15$ & $50$ & 190 & 18.3  \\
    $I_{20}^{50}$ & $20$ & $50$ & 252 & 20.7  \\
    \bottomrule
  \end{tabular}
  \label{tab:single_span_chime}
\end{table}

The single-span AM is an extension of the model proposed in \cite{golik2015convolutional}.
In the first experiment, different kernel sizes $L$ and strides $S$ for $I_S^L$ 
were tested giving the WERs in Table \ref{tab:single_span_chime}. 
The single-span AM gives lower WERs when using smaller kernel sizes, 
with $I^{50}_{10}$ giving a 4.5 \% relative improvement over using the standard kernel 
size of 400 \cite{tuske2018acoustic, sainath2015speaker, golik2015convolutional}.
The input span $T$ makes a noticeable difference to the WERs.
Using $I^{50}_{10}$ as our reference point, a span of $190$ms ($I_{15}^{50}$) 
relatively improves the WER by 5.3 \%. Furthermore, our best performing single-span
AM $I_{15}^{50}$ only gives a slightly worse WER than the baseline $F_{160}^{400}$, and yields a relative 18.4\% improvement 
over the comparable raw waveform system on CHiME4 in \cite{menne18:lafrw}.

%\subsubsection{Multi-Span Acoustic Model}%
%\label{sub:multi_focus_front_end}
In the next experiment,  the proposed 
multi-span structure was investigated
for different constraints on stride and 
kernel size. 
After concatenation, the output vector  $\vect{p}$ of 450-d was fed into the 4L-512d-DNN.
All systems in this section 
use layer-by-layer pre-training by first training one epoch 
on a sub-network where $\vect{p}$ is directly fed into the output layer and then training another epoch 
extending the sub-network with two 512-d hidden DNN layers before the output layer.
We denote the multi-span AM as $M_{S_1,S_2,S_3}^{L_1,L_2,L_3}$ with $L_i$ and $S_i$ giving the
stride and kernel size of the CNN input layer in stream $i \in \{1,2,3\}$.
Table \ref{tab:multi_focus_chime} shows the results. 

\vspace{-2mm}
\begin{table}[th]
  \centering
  \caption{\%WERs with a tg LM and AMs with multiple input streams on CHiME4 dev set.
  Stride combinations $S_1,S_2,S_3$ and kernel size combinations $L_1,L_2,L_3$ are varied.}
  \begin{tabular}{lccccc}
    \toprule
    \multicolumn{1}{l}{ID} & $S$ & $L$ & $T$ & dev \\
    \midrule
    $M_{15,15,15}^{50,100,400}$ & 15 & 50,100,400 & 190-212 & 18.4  \\
    %\cmidrule{1-5}
    $M_{4,9,15}^{50,100,400}$ & 4,9,15 & 50,100,400 & 53-212 & 17.9  \\
    %\cmidrule{1-5}
    $M_{4,9,15}^{50,50,50}$ & 4,9,15 & 50 & 53-190 & 17.1  \\
    \bottomrule
  \end{tabular}
  \label{tab:multi_focus_chime}
\end{table}

For the first system ${M}_{15,15,15}^{50,100,400}$, every input CNN layer 
convolves over the raw waveform signal with the same stride leading to a small range of
input spans 190--212ms. Similar to 
\cite{zhu2016learning}, it was observed that the small kernels mainly 
act as a filter for high frequencies and that the larger kernels 
filter principally lower frequencies, which strongly resembles wavelet filters. 
However, this did not improve the WER over the single-span.
Additionally using different strides in each CNN input layer and therefore 
increasing the range of different spans to $53-212$ms, the system $M_{4,9,15}^{50,100,400}$ yields an improvement 
over the single-span AM.
Finally, all kernels were set to size $50$ and it can be seen  that the system ${M}_{4,9,15}^{50,50,50}$
reduces the WER to 17.1 \% absolute. 
Also, we found that even for a fixed kernel size of $50$, the multi-span AM learns wavelet-like filters by 
setting the weights at the beginning or the end of a kernel to close to zero to effectively 
shorten the kernel size.

\subsection{AMI-IHM}%
\label{sec:ami}
The key results were validated using AMI to 
see how well the model architectures generalize to different datasets.
A baseline based on 40-d FBANK input features was evaluated for comparison\footnote{Considering ${F_{160}^{400}}$ is a small DNN with four 512-d hidden layers and 4k node output layer, and fg LM is used for decoding, its WER is reasonable compared to those in \cite{renals2014}.}.
Table~\ref{tab:multi_span_ami} summarizes the 
results of the key systems ${I}_{10}^{400}, {I}_{10}^{50}$ and ${M}_{4,9,15}^{50,50,50}$ on AMI-IHM. 

\vspace{-2mm}
\begin{table}[th]
  \centering
  \caption{\%WERs with a fg LM and AMs with single and multiple input streams 
  compared to baseline AM based on FBANK on AMI-IHM dev and eval set.}
  \begin{tabular}{ l l c c c }
    \toprule
    \multicolumn{1}{l}{ID} & System & dev & eval \\
    \midrule
    ${F_{160}^{400}}$ & FBANK-DNN &  28.3 & 31.1 \\
    $I_{10}^{400}$ & Single-Span-DNN & 29.1 & 31.9 \\
    $I_{15}^{50}$ & Single-Span-DNN &  28.1 & 30.8 \\
    ${M}_{4,9,15}^{50,50,50}$ & Multi-Span-DNN &  27.2 & 29.3 \\
    \bottomrule
  \end{tabular}
  \label{tab:multi_span_ami}
\end{table}

Table~\ref{tab:multi_span_ami} shows that the single-span AM using raw waveform signal input
gives lower WERs with a smaller kernel size and larger input span also on AMI.
$I_{15}^{50}$ gives a similar WER to the FBANK-DNN AM, while the multi-span 
AM ${M}_{4,9,15}^{50,50,50}$ outperforms the FBANK-DNN AM by a relative WER reduction of 4.8\%.
Comparing ${M}_{4,9,15}^{50,50,50}$ to ${F_{160}^{400}}$ on both AMI and CHiME4 data sets, 
a similar relative WER reduction of 5.5\% is obtained on the CHiME4 dev set. 

\section{Discussion}

\begin{figure}[h]
  \centering
  \includegraphics[width=\linewidth]{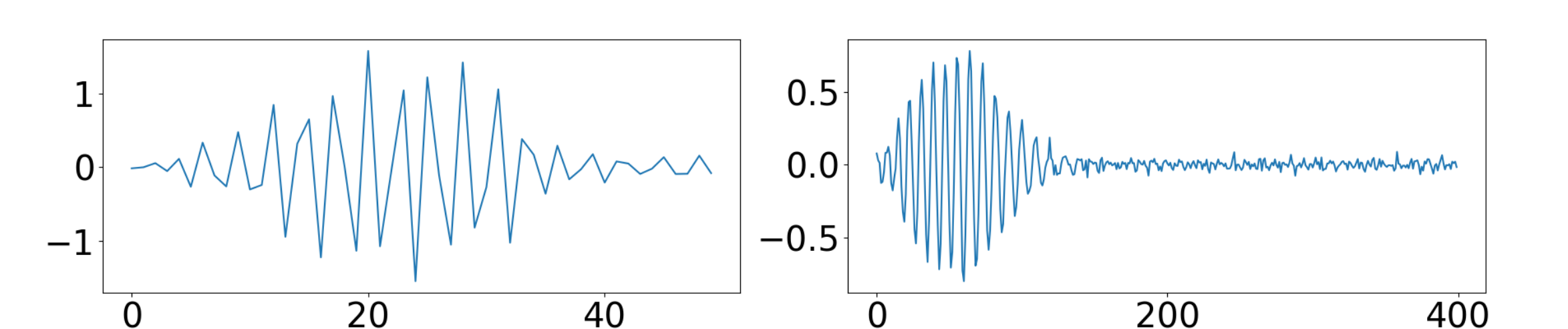}
  \caption{Left: CNN input layer kernel in time domain of size 50 of trained system $I_{15}^{50}$. Right: CNN input layer kernel in time domain of size 400 of trained system $I_{10}^{400}$.}
  \label{fig:filterLen}
\end{figure}

Plotting the input CNN layer kernel weights of the single-span AMs $I_{10}^{400}$ and $I_{15}^{50}$ in the 
frequency domain reveals the typical audiological distributed narrow band pass filters as in 
\cite{sainath2015learning, ghahremani2016acoustic, golik2015convolutional}. 
When plotting the $64$ kernels of size $400$ in the time 
domain, it can be seen that some filter responses are learned only for a small part of the kernel, while the other part is set to zero (\textit{cf.} Fig. \ref{fig:filterLen} right).
While this filter length shortening also happens when a kernel size of $50$ is used, only a much smaller part of the kernel is 
set close to zero (\textit{cf.} Fig. \ref{fig:filterLen} left). This shows that the model automatically learns wavelet-like filters of different time-frequency resolution even for a small fixed kernel size. 

\vspace{-4mm}

\begin{figure}[h]
  \centering
  \includegraphics[width=\linewidth]{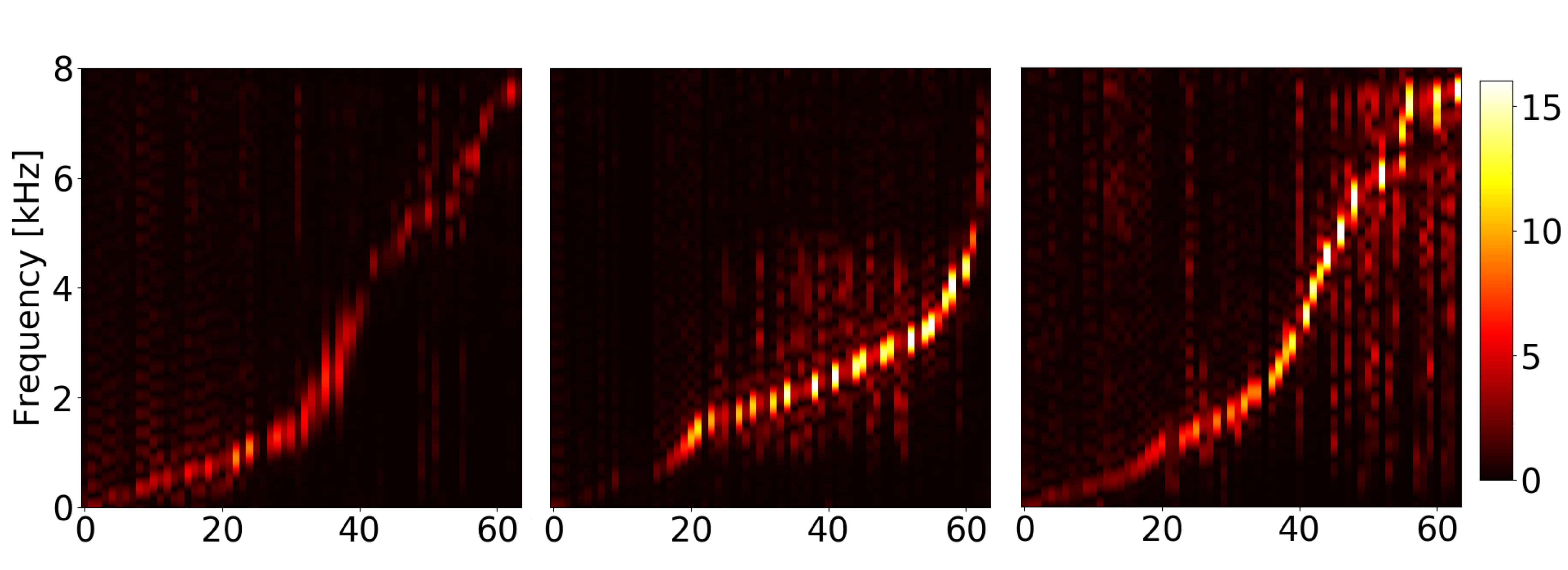}
  \caption{Learned filters of the CNN input layers from our CHiME multi-span AM $M^{50,50,50}_{4,9,15}$ in frequency domain sorted by frequency, which are rather different to the log Mel-filters. Left: Stride $4$, Middle: Stride $9$,
  Right: Stride $15$ .}
  \label{fig:multiSpanFilters}
\end{figure}

In Fig. \ref{fig:multiSpanFilters}, the learned filters 
of the three CNN input layers of $M_{4,9,15}^{50,50,50}$ are smoothed by zero-padding, transformed to the Fourier domain and sorted by frequency. 
It can be seen that the learned filters of the three CNN input layers more or less cover the whole frequency spectrum 
with each filter focusing on a certain area, 
and that they are rather different compared to the log Mel curve used for handcrafted acoustic features.  

\vspace{-4mm}

\section{Conclusions}

We have presented a novel achitecture for acoustic modelling using raw waveform 
input. 
Our model outperforms a conventional DNN-HMM system based on FBANK features 
on the CHiME4 dev set and on the AMI dev and eval sets.
By reducing the kernel size from $400$ to $50$, leaving out any kind of compression layers in the model and 
tuning the input span, we achieved a significant reduction in WER, which questions the usefulness 
of imitating feature extraction pipelines when designing AMs based on raw waveform signal input.
Analysis of the best-performing multi-span AM $M_{4,9,15}^{50,50,50}$ showed that the learned filters are different 
from log-Mel filters in that they do not seem to follow an audiological distribution 
(\textit{cf.} Fig. \ref{fig:multiSpanFilters}).

\section{Acknowledgements}
Thanks to G. Sun for providing the language models for both the AMI-IHM and CHiME4 speech corpus. 
P. von Platen is funded by Studienstiftung des Deutschen Volkes.

\newpage
\section{References}
{\footnotesize 
\renewcommand{\bibsection}{}
%\section{References}

}
\end{document}